\documentclass[aps,groupedaddress,showpacs,letterpaper,twocolumn]{revtex4}

\usepackage{graphicx} 
\usepackage{amsmath}
\usepackage{color}  

\usepackage{graphicx}
\usepackage{amsmath}
\usepackage{color}

\begin{document}

\title{Demonstration of a neutral atom controlled-NOT quantum gate} 

\author{L. Isenhower, E.  Urban, X. L. Zhang, A. T. Gill, T. Henage,  T. A. Johnson$^1$,  T. G. Walker, and M. Saffman}
\affiliation{Department of Physics, University of Wisconsin, 1150 University Avenue, Madison, WI 53706 USA}
\affiliation{1) current address: National Institute of Standards and Technology, Boulder, CO 80305}

\begin{abstract}
We present the first demonstration of a CNOT gate between two individually addressed neutral atoms.  Our implementation of the CNOT uses Rydberg blockade interactions between neutral atoms held in optical traps separated by $ >8~\mu\rm m$. 
Using two different gate protocols we measure  CNOT fidelities of $F=0.73$ and $0.72$ based on truth table probabilities. The gate was used to generate Bell states with fidelity $F=0.48\pm 0.06$. After correcting for atom loss we obtain an {\it a posteriori} entanglement fidelity of $F=0.58.$
\end{abstract}

\maketitle

Any unitary operation can be performed on a  quantum computer equipped with a complete set of universal gates. A complete set of gates can be comprised of single qubit operations together with a two-qubit controlled-NOT (CNOT) gate\cite{Nielsen2000}.
The CNOT gate has been demonstrated in several different physical systems including trapped ions\cite{Monroe1995,Schmidt-Kaler2003}, superconducting circuits\cite{Yamamoto2003,Plantenberg2007}, and linear 
optics\cite{OBrien2003,Pittman2003}. 
Numerous proposals exist for neutral atom quantum gates including
short range dipolar interactions\cite{Brennen1999}, ground state collisions\cite{Jaksch1999}, coupling of atoms to photons\cite{Pellizzari1995}, 
magnetic dipole-dipole interactions\cite{You2000},
gates with delocalized qubits\cite{Mompart2003}, and Rydberg state mediated dipolar interactions\cite{Jaksch2000}. Many particle entanglement mediated by collisions
has been observed in optical lattice based experiments\cite{Mandel2003}, and a $\sqrt{\rm SWAP}$ entangling
operation was performed on many atom pairs in parallel\cite{Anderlini2007}, but a 
quantum gate between  two individually addressed neutral atoms 
has not previously been demonstrated. 

We report here on the demonstration of a two-qubit gate with neutral atoms using Rydberg blockade interactions as proposed in\cite{Jaksch2000}.
The Rydberg approach has a number of attractive features: it does not require cooling of the atoms to the ground state of the confining potentials, it can be operated on $\mu\rm s$ timescales, it does not require precise control of the two-atom interaction strength, and it is not limited to nearest neighbor interactions which is advantageous for scaling to multi-qubit systems\cite{Saffman2008}. Detailed analyses of the Rydberg gate taking into account practical 
experimental conditions\cite{Saffman2005a,Protsenko2002} predict that gate errors at the level of  
$F\sim10^{-3}$ are possible. We present here  initial demonstrations of  Rydberg mediated CNOT 
gates with fidelities based on truth table probabilities of 
$F=0.73,0.72$ using two different protocols. 
The coherence of the gate is shown by measuring coherent oscillations of the output states, with a
conditional  phase that is dependent on the presence or absence of a  two-atom Rydberg interaction. 
Using superposition input states the gates were used to generate Bell states with fidelity $F=0.48$ which suggests the actual gate fidelity lies between $0.48$ and $0.73$.

Our implementation of the CNOT gate builds on  earlier  demonstrations of single qubit rotations using two-photon stimulated 
Raman pulses\cite{Yavuz2006}, coherent excitation of Rydberg states\cite{Johnson2008}, and Rydberg blockade\cite{Urban2009,Gaetan2009}. 
The experimental apparatus and procedures used for excitation of Rydberg states are similar to that described in \cite{Urban2009}. As shown there, excitation of a control atom to a Rydberg level with principal quantum number $n=90$ prevents subsequent excitation of a target atom in a neighboring site separated by $R=10~\mu\rm m$. Excitation and de-excitation of the target atom corresponds to a $2\pi$ rotation of an
effective spin $1/2$ which therefore imparts a $\pi$ phase shift to the wavefunction of the target atom. If the control atom blocks the target excitation then the rotation does not occur and there is no phase shift of the target wavefunction. The result is a $C_Z$ controlled  phase operation.

\begin{figure}[!t]
\centering
\includegraphics[width=8.6cm]{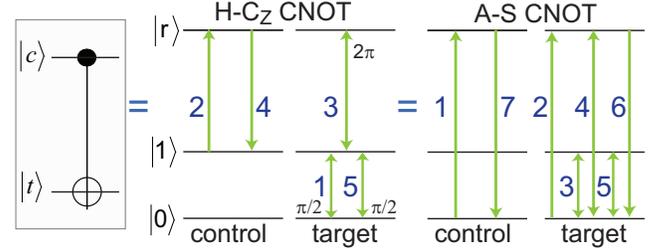}
\vspace{-.5cm}
\caption{(color online) 
CNOT gate protocols using Hadamard - C$_{Z}$ (center) and controlled amplitude swap (right). 
All pulses are $\pi$ pulses except where indicated otherwise.}
\label{fig.1}
\end{figure}

There are several possible ways to convert the Rydberg blockade 
operation into a full  CNOT gate. A standard approach\cite{Nielsen2000} shown in Fig. 1. is to  perform Hadamard rotations on the target qubit before and after the
controlled phase which immediately generates what we will refer to as a H-C$_Z$  CNOT. 
An alternative is to implement a controlled amplitude swap (A-S CNOT), as was originally proposed in  the context of rare earth doped crystals\cite{Ohlsson2002}. As seen in Fig. 1 (right), when the control atom is initially in state $|0\rangle$ it is excited to 
the Rydberg level $|r\rangle$ by pulse 1 and the blockade interaction prevents Rydberg pulses $2,4,6$ on the target atom from having any effect. The final pulse $7$ returns the control atom to the ground state. When the control atom is initially in state $|1\rangle$ pulses $1$ and $7$ are detuned and have no effect, while pulses $2-6$ swap the amplitudes of $|0\rangle$ and $|1\rangle$. This corresponds to a standard CNOT apart from 
a single qubit phase that can be corrected.  This approach 
to generating the CNOT where we have a conditional state transfer, instead of the more usual conditional phase, can be further 
 generalized to efficiently generate  many-atom entanglement\cite{Saffman2009b,Muller2009}.

\begin{figure}[!t]
\includegraphics[width=8.5cm]{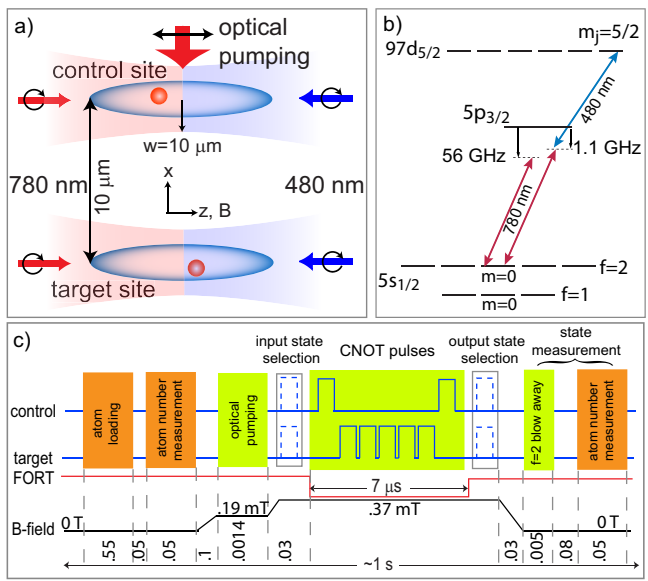}
\vspace{-.3cm}
\caption{\label{fig.exp}(color online) a) Experimental geometry, b) laser excitation frequencies, and c) CNOT sequence.  See text for details.}
\end{figure}

Our experimental approach shown in Fig. \ref{fig.exp} follows that  described more fully in \cite{Urban2009} and the associated
supplementary information. Single $^{87}$Rb atoms are localized in far off resonance traps (FORTs) created by focusing a laser propagating along $+z$ with  wavelength $\lambda=1064~\rm nm$, to spots with  waists ($1/e^2$ intensity radius) $w=3.0~\mu\rm m$. 
The resulting traps  have  a potential depth of $U/k_B=5.1~\rm mK.$ The trapped atoms have a measured temperature of  $T \simeq 200-250 ~\mu\rm K. $
The  position probability distributions for each atom are approximately  Gaussian with $\sigma_{x,y}\sim0.3~\mu\rm m$ and $\sigma_z\sim4~\mu\rm m$. A bias magnetic field
is applied along $z$, while the sites are separated by a distance of about $10~\mu\rm m$ along $x$.  
After optical trapping the atoms are optically pumped into $|f=2,m_f=0\rangle$ using $\pi$ polarized light propagating along $-x$ tuned to the 
$|5s_{1/2},f=2\rangle \rightarrow |5p_{3/2},f=2\rangle$
and 
$|5s_{1/2},f=1\rangle \rightarrow |5p_{3/2},f=2\rangle$ transitions. Atom measurements are performed by collecting resonance fluorescence on a cooled CCD camera, and comparing the integrated number of counts in a region of interest
with predetermined thresholds indicating the presence or absence of a  single atom\cite{Urban2009}.

Laser beams for ground state rotations and Rydberg excitation are focused to 
near circular waists of $w\sim 10 ~\mu\rm m$(see \cite{Urban2009} for the measured ellipticity of the beams) and propagate along $+z$ (780 nm) and $-z$ (480 nm). The beams can be 
switched to address either site as described in \cite{Urban2009}.
Ground state Rabi pulses are generated by focusing a
$\sigma_+$ polarized  780 nm laser with frequency components separated by $6.8~\rm  GHz$ and detuned by 56 GHz to the 
red of the $|5s_{1/2}\rangle-|5p_{3/2}\rangle$ transition (see \cite{Yavuz2006} for further details).
Typical total power in the two Raman sidebands is $\sim 85~\mu\rm W$ and we achieve $\pi$ pulse times of $\sim 600 ~\rm ns.$

Rydberg excitation uses $\sigma_+$ polarized 780 and 480 nm beams tuned for resonant excitation 
of the Rydberg state $|97d_{5/2},m_j=5/2\rangle.$  
A bias field of $0.37 ~\rm mT$ is used to shift the $m_j=3/2$ state by  $-6.2~\rm MHz$ relative to the
$m_j=5/2$ state   so that $m_j=3/2$ is not significantly populated by the Rydberg lasers.    The 780 nm beam is tuned about $1.1 ~\rm GHz$ to the red of the $|5s_{1/2},f=2\rangle \rightarrow |5p_{3/2},f=3\rangle$ transition. Typical beam 
powers are $2.3~\mu\rm W$ at 780 nm and $12~\rm mW$ at 480 nm giving Rydberg $\pi$ pulse times of 
$\sim 750~\rm ns$\cite{cnotnote}. The target atom pulses 2-6 in Fig. 1 (right) are given by the sequence $R_r(\pi)R_g(\pi)R_r(\pi)R_g(\pi)R_r(\pi)$ where $R_{g/r}(\theta)$ are pulses of area $\theta$ between
$|0\rangle \leftrightarrow|1\rangle / |0\rangle \leftrightarrow |r\rangle.$

The experimental sequence for demonstrating the CNOT is shown in Fig. 2. We start by loading one atom into each optical trap which is verified by the first atom number measurement. Both atoms are then optically pumped into $|f=2,m_f=0\rangle$ and ground state $\pi$ pulses are applied to either or both of the atoms to generate any of the four computational basis states. 
We then turn off the optical trapping potentials, apply the CNOT pulses of Fig. 1, and restore the optical traps. Ground state $\pi$ pulses are then applied to either or both atoms to  select one of the four possible output states, atoms left in state $|f=2\rangle$ are removed from the traps with unbalanced radiation pressure (blow away light), and a measurement is made to determine if the selected output state is present.

\begin{figure}[!t]
\centering
\includegraphics[width=8.7cm]{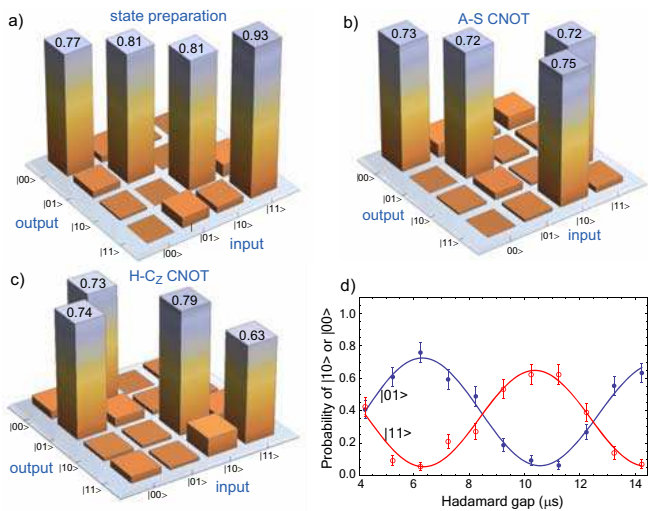}
\vspace{-.9cm}
\caption{\label{fig.cnotdata}(color online) Measured probabilities for a) state preparation, b) A-S CNOT, c) H-C$_Z$ CNOT and d) output states of the  H-C$_Z$ CNOT under variation of the relative phase of the $\pi/2$ pulses. 
The reported matrices are based on an average of at least 100 data points for each matrix element and the error bars are $\pm 1$ standard deviation.  
}
\end{figure}

The time needed to apply the CNOT pulses is about  $7~\mu\rm s$, 
while the entire cycle time is approximately $1~\rm  s$. The difference is primarily due to a $0.55~\rm  s$ atom loading phase and the time needed to 
turn on and off  bias magnetic fields which are used during 
optical pumping and Rydberg excitation. 
The experiments are performed in a vacuum chamber with a pressure of about $2\times 10^{-7}~\rm Pa$. Collisions of the trapped atoms with
hot background atoms result in a finite lifetime of the trapped atoms with an exponential time constant measured to be about $3~\rm  s.$ 
We therefore expect a collisional loss during the $0.3~\rm  s$ gap between the first and second measurements of about  10\%, which is confirmed by measurements. In addition there is a $\sim 5\%$ loss probability due to turning the trapping potential off for $7~\mu\rm s$. These losses occur independent of the CNOT gate operation and we therefore normalize all two-atom data reported below by a factor of $1/.85^2$ to compensate for this loss. The reported data were obtained over a period of several months during which the two-atom loss factor
varied by no more than $\pm 2\%$. 
We expect that future experiments with better vacuum, colder atoms, and shorter gap time will remove the need for this correction factor. 

It is important to emphasize that our use of selection pulses provides a positive identification of all output states and we do not simply assume that a low photoelectron signal corresponds to an atom in $|0\rangle$ before application of the blow away light. This is important because of the nonzero probability of atom loss mentioned above during 
each experimental sequence. 
For example, if we wish to verify the presence of the state $|01\rangle$ we apply a $\pi$ selection pulse to the control atom and no selection pulse to the target atom. We then apply blow away light to both atoms and measure if there is still an atom in both sites. A positive signal for both atoms (number of photoelectron counts between the precalibrated single atom and two atom limits)
signals the presence of $|01\rangle$. Changing the selection pulses 
identifies the presence of any of the four possible states.


Following the above procedures we have obtained the CNOT truth tables
shown in Fig. 3. For the H-C$_Z$ CNOT states $|0(1)\rangle$ are $|f=1(2),m_f=0\rangle$ and for the A-S CNOT
$|0(1)\rangle$ are $|f=2(1),m_f=0\rangle.$
In Fig. \ref{fig.cnotdata}a) we show the fidelity of our state preparation, which is obtained using the sequence of Fig. \ref{fig.exp} but without 
applying the CNOT pulses. The computational basis states are prepared with an average fidelity of $F=0.83$.    The measured probability matrices for the state preparation and the CNOT gates are 
\vspace{-.1cm}
$$
U_{\rm prep}=\begin{pmatrix}
0.77 &0.04 &0.01& 0.0\cr
0.04 &0.81 &0.0 &0.0\cr
0.02 &0.0 &0.81 &0.08\cr
0.0 &0.07 &0.04 &0.93
\end{pmatrix},\vspace{-.3cm}
$$
and
\vspace{-.1cm}
\begin{eqnarray}
\begin{pmatrix}
0.73 &0.08 &0.02& 0.08\cr
0.0 &0.72 &0.02 &0.03\cr
0.01 &0.04 &0.02 &0.72\cr
0.0 &0.02 &0.75 &0.03
\end{pmatrix}_{\hspace{-.2cm}A-S}
\hspace{-.2cm},\hspace{-.01cm}
\begin{pmatrix}
0.05 &0.73 &0.0& 0.02\cr
0.74 &0.06 &0.02 &0.03\cr
0.02 &0.02 &0.79 &0.06\cr
0.04 &0.02 &0.12 &0.63
\end{pmatrix}_{\hspace{-.2cm}H-C_{Z}}.
\nonumber
\end{eqnarray}
We believe that the finite fidelity of state preparation can be largely attributed to imperfect optical pumping, and small drifts in our preparation and analysis laser pulses. 
The H-C$_Z$ CNOT was obtained using $\pi/2$ pulses that were $\pi$ out of phase which inverts the gate matrix relative to that seen for the A-S CNOT.

The fidelity of transferring the input states to the correct output states is  
$F=\frac{1}{4}{\rm Tr}[|U_{\rm ideal}^T| U_{\rm CNOT}]=0.73,0.72$ for the A-S and H-C$_Z$ CNOT gates. We note that the average ratio of the ``high" truth table elements to the ``low" elements is about $25:1$ which  would imply a CNOT fidelity above 0.95. The lower value of the observed fidelity can be attributed to imperfect  state preparation, 
errors in the applied pulses, and a small amount of blockade leakage.
The calculated interaction strength\cite{Walker2008} for atoms separated by $10.2~\mu\rm m$ along $x$ including the effect of the bias magnetic field,  is $B/2\pi=9.3~\rm MHz.$ With a Rydberg  excitation Rabi frequency of $\Omega/2\pi = 0.67~\rm MHz$ this implies a residual double excitation probability 
due to imperfect blockade of $P_2\simeq \Omega^2/(2 B^2) = 2.6\times 10^{-3}.$ At a temperature of 
$200~\mu\rm K$ we expect two atom separations along $z$ extending out to   $\Delta z\sim 10~\mu\rm m$ to occur with $\sim 10\%$ probability. Averaging over the thermal distribution of atomic separations implies a double excitation probability of $\bar P_2\sim 0.1.$   Pulse area and blockade errors imply a  non-zero amplitude for either atom to be in a Rydberg state at the end of the gate. In such cases the Rydberg atom is photoionized when the optical trapping potentials are restored which results in atom loss. This is evident in that the average probability sum from each column of Fig. 3 is $0.90$ for the state preparation but only $0.82,0.84$ for the CNOT gates. 

The intrinsic coherence of the H-C$_Z$ CNOT gate is seen in Fig. \ref{fig.cnotdata}d) where the output probabilities are shown with a varying gap between  pulses 4 and 5 in Fig. \ref{fig.1} for  input states $|ct\rangle=|01\rangle$ and $|11\rangle$. 
Varying the gap time changes the relative phase since our ground state Raman beams are two-photon detuned from the $|1,0\rangle - |2,0\rangle$ transition, to account for the Raman light induced AC Stark shift\cite{Yavuz2006}. The coherent  oscillations of the output state curves are $\pi$ out of phase for the control atom in state $|0\rangle$ or $|1\rangle$, corresponding to the {\it conditional} $\pi$ phase shift from Rydberg blockade.

\begin{figure}[!t]
\centering
\includegraphics[width=8.6cm]{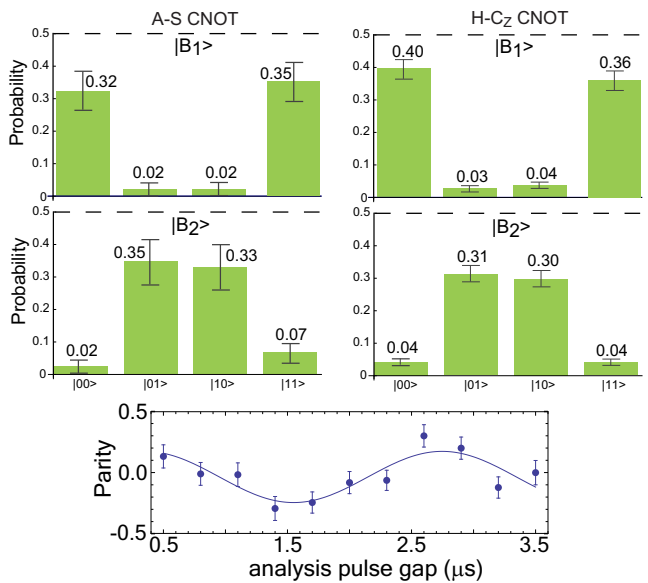}
\vspace{-.7cm}
\caption{\label{fig.belldata}(color online) 
Measured probabilities for preparation of Bell states $|B_1\rangle, |B_2\rangle$ using A-S CNOT (left) and H-C$_Z$ CNOT (right).  The parity oscillation data was obtained from $|B_1\rangle$  with the H-C$_Z$ CNOT.  
}
\end{figure}

The importance of the CNOT gate stems in  part from its ability  to deterministically create entangled states\cite{Turchette1998}. To investigate this we used $\pi/2$ pulses on the control atom to prepare the input states
$|ct\rangle=\frac{1}{\sqrt{2}}(|0\rangle+i|1\rangle)|0\rangle$ and 
$|ct\rangle=\frac{1}{\sqrt{2}}(|0\rangle+i|1\rangle)|1\rangle$.
Applying the CNOT to these states creates two of the Bell 
states $|B_1\rangle=\frac{1}{\sqrt{2}}(|00\rangle+|11\rangle)$
and
$|B_2\rangle=\frac{1}{\sqrt{2}}(|01\rangle+|10\rangle)$.
The measured probabilities for these output states are shown in Fig. 4. 
 
In order to verify entanglement of the Bell states we measured the parity signal $P=P_{00}+P_{11}-P_{01}-P_{10}$
after applying delayed $\pi/2$ analysis pulses to both atoms with a variable phase $\phi$\cite{Turchette1998}.
A short calculation shows that the parity signal varies as $P=2 {\rm Re}(C_2)-2 |C_1| \cos(2\phi+\xi)$ where
$C_2$ is the coherence between states $|01\rangle$ and $|10\rangle,$ and  $C_1=|C_1|e^{\imath\xi}$ is the coherence between states $|00\rangle$ and $|11\rangle$. A curve fit yields ${\rm Re}(C_2)=-.02,$
and $|C_1|=0.10.$ The fidelity of entanglement of the Bell state $|B_1\rangle$ can be quantified by\cite{Sackett2000} $F=\frac{1}{2}(P_{00}+P_{11})+|C_1|.$ States with $0.5< F \le 1$ are entangled. The data in Fig. 
\ref{fig.belldata} yield $F=0.48\pm.06$ which is just under the threshold of $F=0.5$ for entanglement. The trace of the density matrix for the state $|B_1\rangle$ in Fig.\ref{fig.belldata} is ${\rm Tr}[\rho]=0.83$ due to atom loss as discussed above. Dividing by ${\rm Tr}[\rho]$ to correct for atom loss implies that the atom pairs which remain after the gate are entangled with fidelity 
$F=0.58.$ 
This form of {\it a posteriori} entanglement\cite{vanEnk2007} is useful 
for further quantum processing since it could in principle be converted into genuine heralded entanglement by running the gate on two pairs of atoms followed by entanglement swapping\cite{Zukowski1993}. 
A complementary method resulting in entanglement of the surviving  atom pairs with $F=0.75$ has independently been demonstrated by Wilk, et al. \cite{Wilk2009}.

In conclusion we have  presented the first realization of a CNOT gate between two individually addressed neutral atoms.
Coherent oscillations of the output state demonstrate the effect of the conditional phase from the two-atom Rydberg interaction, and we have used the gate to generate states on demand near the threshold of entanglement. Correcting for atom losses we have entanglement with $F=0.58$ between pairs of atoms remaining after the gate. Our
results are obtained with atoms at $T\sim 200~\mu\rm K$, which confirms that cooling to the motional ground state 
is not required for the blockade gate. Additional cooling will, however, improve the spatial localization of the atoms and we anticipate that better cooling,  as well as other improvements,  will lead to higher gate 
fidelities in the future. 
The gate is performed with atoms that are  separated by more than $8~\mu\rm m$. The use of more tightly confining optical traps and more tightly focused optical beams  will allow these experiments to be performed with interatomic spacings as small as a few microns which implies the feasibility of scaling these results to multiparticle entanglement of tens of atoms.

This work was supported by  NSF grant PHY-0653408  and ARO/IARPA under contract W911NF-05-1-0492.



\begin{thebibliography}{32}
\bibitem[{cno()}]{cnotnote}
\bibinfo{note}{The parameters specified in the text are for the A-S {CNOT}. For
  the H-C$_Z$ CNOT we used $w\sim5~\mu\rm m$, one-photon detuning of the Raman
  laser of 100 GHz and detuning of the Rydberg lasers from the $5p_{3/2}$ state
  of 2.1 GHz.}
\end{thebibliography}

\end{document}